\documentclass[12pt]{article}

\usepackage{amsmath,amssymb}
\usepackage{cite}
\usepackage{graphicx}
\usepackage[utf8]{inputenc}

\begin{document}

\begin{titlepage}
\begin{center}

\vspace{-0.4in}

{\large \bf Directed Polymers and Interfaces in Disordered Media}\\

\vspace{.25in}

%%%%%%%%%%%%%%%%%%%%%%%%%%%%%%%%%%%%%%%%%%%%%%%%%%%%%%%%%%%%%%%%%%%%%%%%
{\large \em {Róbinson J. Acosta Diaz \footnotemark[1]}}\\
\vspace{.08in}

Instituto de F\'{i}sica, Universidade Federal do Rio de Janeiro, \\
\vspace{.06in}
Av. Athos da Silveira Ramos 149,  Rio de Janeiro 21941-972, Brazil\\
%%%%%%%%%%%%%%%%%%%%%%%%%%%%%%%%%%%%%%%%%%%%%%%%%%%%%%%%%%%%%%%%%%%%%%%%

\vspace{.25in}

%%%%%%%%%%%%%%%%%%%%%%%%%%%%%%%%%%%%%%%%%%%%%%%%%%%%%%%%%%%%%%%%%%%%%%%%
{\large\em Christian D. Rodríguez-Camargo \footnotemark[2]}\\
\vspace{.08in}

Departamento de Ingeniería Industrial, Corporación Universitaria Minuto de Dios, \\
\vspace{.06in}
Carrera 73A No. 81B-70, Bogotá 111021, Colombia

\vspace{.14in}

Facultad de Ingeniería, Universidad Nacional de Colombia—Sede Bogotá, \\
\vspace{.06in}
Calle 44 No. 45-67, Bogotá 111321, Colombia
%%%%%%%%%%%%%%%%%%%%%%%%%%%%%%%%%%%%%%%%%%%%%%%%%%%%%%%%%%%%%%%%%%%%%%%%

\vspace{.25in}

%%%%%%%%%%%%%%%%%%%%%%%%%%%%%%%%%%%%%%%%%%%%%%%%%%%%%%%%%%%%%%%%%%%%%%%%
{\large\em  Nami F. Svaiter \footnotemark[3]} \\
\vspace{.08in}

Centro Brasileiro de Pesquisas F\'\i sicas, \\
\vspace{.06in}
Rua Dr. Xavier Sigaud 150, Rio de Janeiro 22290-180, Brazil\\
%%%%%%%%%%%%%%%%%%%%%%%%%%%%%%%%%%%%%%%%%%%%%%%%%%%%%%%%%%%%%%%%%%%%%%%%

\vspace{.3in}

\subsection*{\\Abstract}
\end{center}

\baselineskip .22in
\noindent
We consider field theory formulation for directed polymers and interfaces in the presence of 
quenched disorder. We write a series representation for the averaged free energy, where~ all the integer 
moments of the partition function of the model contribute. The structure of field space is analysed for 
polymers and interfaces at finite temperature using the saddle-point equations derived from each integer 
moments of the partition function. For the case of an interface we obtain the wandering exponent 
$\xi=(4-d)/2$, also obtained by the conventional replica method for the replica symmetric scenario.

\bigskip
%\vspace{.06in}

\noindent
{\sc Keywords:} disordered systems; free energy; wandering exponent
\vspace{.06in}
 
\footnotetext[1]{e-mail:\,\,racosta@if.ufrj.br}
\footnotetext[2]{e-mail:\,\,cdrodriguezc@unal.edu.co}
\footnotetext[3]{e-mail:\,\,nfuxsvai@cbpf.br}

%\vspace{.09in}
%\noindent
%PACS numbers: 05.20.-y,\,75.10.Nr

\end{titlepage}
\newpage
\baselineskip .22in

\section{Introduction}

The statistical mechanics of random surfaces and membranes, or more generally of extended objects, has 
been widely discussed in the literature, see, e.g., \cite{membranas}. One of the simplest example of a 
tethered surfaces are the polymers \cite{halpin, fredrickson, kacokatso}. Certain models of polymers can 
be discussed by a classical field theory \cite{deGennes, des, witten}. Directed polymers in the presence 
of a quenched random potential describe, for~ example, the behavior of a linear  elastic objects with no 
self-intersections in a porous medium and also the polymer behavior in poor solvents \cite{vilgis,craig}. 
The generalization to more complex extended objects is straightforward 
\cite{duo,du0,du1,nat,mezard-parisi1, mezard-parisi3}. One can also consider a $d$-dimensional manifold 
with internal points $x \in \mathbb {R}^{d}$, embedded in an external $D$-dimensional space with position 
vector $\vec{r}\,(x) \in \mathbb {R}^{D}$, where~ $D=d+N$. For~ oriented manifolds, the set of $N$ 
transverse coordinates describe the fields of the model.  The~ $N=1$ case describes an interface in a 
quenched random potential. In these systems with disorder, two averages must be performed. The average of 
a thermal ensemble using the Boltzmann weight  and also the average over all the realizations of the 
disordered variables.

For quenched disorder, one is mainly interested in averaging the free energy over the disorder, which 
amounts to averaging the logarithm of the partition function. In the field theory of random manifolds, a 
technique that has been used in order to compute the average free energy is the replica method 
\cite{edwards,livro1,parisi1,livro3, livro4,livromez}. It is known that there are criticisms concerning 
the fundamental mathematics behind such method \cite{sourlas,FTdisorsys,errado0,errado1,errado2,errado3}. 
Nevertheless this procedure has succeed in describing polymers and membranes in a random media. A nice 
discussion about the $n\rightarrow0$ limit can be found in Ref. \cite{ma,gaspari} The main mathematical 
problem of the replica trick is that it is not possible to interpret the above discussed limit as an 
analytical continuation procedure. The aim of this paper is to present an alternative to the replica 
trick.  We study directed polymers and fluctuating interfaces in quenched random potentials, where we 
compute the wandering exponent in a generic $d$-dimensional manifold. We obtain that the wandering 
exponent is given by $\xi=\frac{4-d}{2}$ as was discussed in Ref. \cite{mezard-parisi2}, for the replica 
symmetric structure of the correlation functions.   

A new method to calculate the average free energy of systems defined in the continuum with quenched 
disorder was presented in Refs. \cite{svaiter1,svaiter2}. An application of such procedure was presented 
in Ref. \cite{novo,gastao} where a Landau-Ginzburg model with a disorder field linearly and quadratically 
coupled with the order parameter was discussed. The static version of a non-relativistic field theory with 
a complex field was investigated in \cite{novo1}. There, an interacting boson system below the critical 
condensation temperature was studied. It was discussed the effects of quenched disorder in a dilute Bose-
Einstein condensate confined in a hard walls trap. Using the disordered Gross-Pitaevskii functional, a 
representation  for the quenched free energy as a series of integer moments of the partition function was 
obtained, where  positive and negative disorder-dependent effective coupling constants appear in the 
integer moments. The combined contributions of effects due to boundary conditions and disorder in the 
weakly disordered condensate was analysed, and the ground state renormalized density profile of the 
condensate was presented. This new technique was also used to discuss fluctuations of the Hawking 
temperature in an Euclidean Schwarzschild manifold \cite{novo2}.   

The organization of this paper is as follows. In Section \ref{sec:III}, we discuss directed polymers in 
the presence of a quenched disorder. In Section \ref{sec:IV}, the average free energy associated with a 
manifold in the presence of a quenched disorder is presented. Conclusions are given in Section 
\ref{sec:V}. We use $\hbar=c=k_{B}=1$.

\section{The Field Theory in $D=1$ with Quenched Disorder} \label{sec:III}
Let us consider a directed polymer of length $L$, where for simplicity we assume that the 
displacements of the polymer can occur in one direction. In the continuum approximation, 
the~ Hamiltonian of the directed polymer can be written as

\begin{equation}
 H(\varphi,v)=\int_{0}^{L}dx\Biggl[\frac{c}{2}\biggl(\frac{d\varphi}{dx}\biggr)^{2}
                                                  +v\bigl(\varphi(x),x\bigr)\Biggr],
\label{hamiltonian}
\end{equation} 
where $c$ is the linear tension of the polymer, $x$ is the longitudinal coordinate $(0\leq x \leq L)$,
and $\varphi(x)$ is the transverse displacement of the polymer with respect to the straight line. Finally, 
$v(\varphi(x),x)$ is the quenched disordered potential of the model 
\cite{larkin,huse,kardar,garel,digkb,dggb}. There are different proposed probability distributions 
associated with the disorder. A widely used probability distribution is the Gaussian (normal) distribution. 
We will take $v(\varphi(x),x)$ to be a Gaussian random variable which has zero mean and it is 
delta-correlated in the transverse direction. Therefore

\begin{equation}
\mathbb{E}\bigl(v(\varphi,x)\bigr)=0
\label{correlation1}
\end{equation} 
and

\begin{equation}
\mathbb{E}\bigl(v(\varphi,x)v(\varphi',x')\bigr)=2V(\varphi-\varphi')\delta(x-x'),
\label{correlation2}
\end{equation} 
where $\mathbb{E}(...)$ means the average over all realizations of the quenched random potential and 
$V(\varphi-\varphi')$ stands for the correlation function of the model. The scaling relation defines the 
wandering exponent. We have 

\begin{equation}
\mathbb{E}\bigl[\langle(\varphi(L)-\varphi(0))^{2}\rangle\bigr]\propto L^{2\xi},
\label{squared-displacement}
\end{equation}
where $\langle...\rangle$ is a thermal average, i.e., the configurational average of the Boltzmann weight. 
The~ quantity $\mathbb{E}\bigl[\langle(\varphi(L)-\varphi(0))^{2}\rangle\bigr]$ is the polymer mean squared 
displacement $\varphi$ with length $L$ where $\xi$ is the wandering exponent. The partition function of 
the model can be written as 

\begin{equation}
 Z(L,y;v)=\int_{\varphi(0)=0}^{\varphi(L)=y}[d \varphi]\exp\bigl(-\beta H(v,\varphi)\bigr), 
\label{partition-function}
\end{equation} 
where $[d \varphi]$ is a formal functional measure. The average free energy is defined as

\begin{equation}
 F_{q}=-\frac{1}{\beta}\int[dv]P(v)\log Z(L,y;v),
\end{equation}      
where $[dv]\,P(v)$ is the probability distribution associated with the disorder. To obtain the 
average free energy of the model we can use the replica method. For an application of this method to study 
finite size effects in the random field Ising model see the Ref. \cite{periodic}.

Inspired in the usual situation where one defines  zeta functions in terms of countable collections of 
numbers \cite{riem2,primes0,primes,voros} and also defining the zero point energy of quantum fields in the 
presence of boundaries \cite{nami1,nami2,nami3}, we define the distributional zeta function as

\begin{equation}
\Phi(s)=\int [dv]P(v)\frac{1}{Z(L,y;v)^{s}},
\label{pro1}
\end{equation}
for $s\in \mathbb{C}$, this function being defined in open connected subset of the complex plane. The 
average free energy can be computed using

\begin{align}
F_{q}=\left.\frac{1}{\beta}\frac{d}{ds}\Phi(s)\right|_{s\rightarrow0^{\,+}}=
\left.\frac{1}{\beta}\int [dv]P(v)\frac{d}{ds}\frac{1}{Z(L,y;v)^{s}}\right|_{s\rightarrow0^{\,+}}.
\label{sa22}
\end{align}

{
Next, one can write $1/Z^{s}$ using the Euler integral representation for the Gamma function. Breaking~ 
this integral representation into two integrals, one from zero to $a$ and another from $a$ to infinity, 
where $a$ is an arbitrary real number, and expanding the exponential in power series in the first 
integral, one can write that the average free energy is} \cite{svaiter1,svaiter2} 

\begin{equation}
F_q=\frac{1}{\beta}\sum_{k=1}^\infty \frac{(-1)^{k}a^{k}}{k!k}\,\mathbb{E}({Z^{\,k}})
+\frac{1}{\beta}\bigr(\ln a+\gamma\bigl)-\frac{1}{\beta}R(a),
\label{serie}
\end{equation}
where $\gamma$ is the Euler-Mascheroni constant and 

\begin{equation}
R(a)=-\int  [dv]P(v)\int_{a}^{\infty}\,\dfrac{dt}{t}\, e^{-Z(L,\,y;\,v)\,t}.
\label{m24}
\end{equation}

For large $a$ it is possible to work only with the series contributions. Next,  we absorb the 
dimensionless parameter $a$ in the functional integral measure. To proceed, let $Z^{\,k}$ be the $k$-th 
power of the partition function, for $k$ integer. In this case, we have a perturbative expansion of the 
average free energy given by Equation (\ref{serie}). The $k$-th power of the partition function $Z^{\,k}$ can 
be written as

\begin{equation}
(Z(L,y;v))^{k}=\,\prod_{i=1}^{k}\int_{\varphi_{i}(0)=0}^{\varphi_{i}(L)=y}[d\varphi_{i}]\,
                                     \exp\biggl(-\beta\sum_{a=1}^{k}H(\varphi_{a},v)\biggr).                             
\label{u1}
\end{equation}

Averaging $(Z(L,y;v))^{k}$ over the disorder we obtain that the $k$-th moment of the partition function is 
given by

\begin{equation}
\mathbb{E}(Z^{k})=\prod_{i=1}^{k}\int_{\varphi_{i}(0)=0}^{\varphi_{i}(L)=y}[d\varphi_{i}]\,
                                              \exp\bigl(-\beta H_{eff}(\varphi_{i},k)\bigr),
\label{aa11}
\end{equation}
where the effective Hamiltonian $H_{eff}(\varphi_{i},k)$ is

\begin{equation}
 H_{eff}(\varphi_{i},k)=\int_{0}^{L}dx\Biggl[\frac{c}{2}\sum_{i=1}^{k}
                        \biggl(\frac{d\varphi_{i}}{dx}\biggr)^{\,2}-\beta\sum_{i,j=1}^{k}V
                                                      \bigl(\varphi_{i}(x)-\varphi_{j}(x)\bigr)\Bigg].
\label{heff1}
\end{equation}

{
We would like to stress that the method used in this paper, contrary to the replica method neither 
involves derivatives of the integer moments of the partition function, nor the extension to this 
derivative to non-integer values of $k$.} Now, we have to discuss the quantity $V(\varphi(x)-\varphi'(x))$ 
defined in Equation (\ref{correlation2}). It is well know that the delta correlated potential 
$V(\varphi)=u\delta(\varphi)$ maps the replicated problem to interacting quantum bosons.Using that 
$V(\varphi-\varphi')$ is given by

\begin{equation}
 V(\varphi-\varphi')=V_{0}-\frac{1}{2}u(\varphi-\varphi')^{2},
\label{app}
\end{equation}  
permits an entire analysis via replicas \cite{dggb}. Since we  are interested in a soluble model, we 
assume that $V(\varphi-\varphi')$ is given by Equation (\ref{app}). After integrating by parts, we can write 
that $H_{eff}=H_{eff}^{(1)}+H_{eff}^{(2)}$,~ where

\begin{equation}
 H_{eff}^{(1)}(\varphi_{i},k)=\frac{1}{2}\int_{0}^{L}dx\sum_{i,j=1}^{k}\varphi_{i}(x)
     \Biggl[\biggl(-c\frac{d^{\,2}}{dx^{2}}+\beta u\biggr)\delta_{ij}-\beta u\Bigg]\varphi_{j}(x),
\label{heff2}
\end{equation}
and

\begin{equation}
 H_{eff}^{(2)}=\int_{0}^{L}dx\,\beta V_{0}.
\end{equation}

Studying the replica field theory for the problem of fluctuating manifold in a quenched random potential, 
M\'ezard and Parisi and others introduced a mass term in the effective Hamiltonian in order to regularize 
the model \cite{mezard-parisi2,doussal}. Indeed, in the high temperature limit, i.e., 
$\beta\rightarrow 0$ the operator $(-\frac{c}{2}\frac{d^{\,2}}{dx^{2}}+\beta u)$ has the zero eigenvalue 
and therefore is not invertible. To circumvent this problem, we~ are following the same idea introducing 
the term

\begin{equation*}
\frac{1}{2}\int_{0}^{L}\,dx\bigl[\varphi_{i}(x)\omega^{2}\delta_{ij}\varphi_{j}(x)\bigr],
\end{equation*}
in the effective Hamiltonian $H_{eff}$. Neglecting $H_{eff}^{(2)}$, we have

\begin{equation}
  H_{eff}(\varphi_{i},k;\omega)=\frac{1}{2}\int_{0}^{L}dx\sum_{i,j=1}^{k}
	\varphi_{i}(x)\Biggl[\biggl(-c\frac{d^{\,2}}{dx^{2}}+\omega^{2}+\beta u\biggr)\delta_{ij}
	  -\beta u\Bigg]\varphi_{j}(x).
\end{equation} 

In the limit where $\omega\rightarrow\, 0$ we recover the polymer field theory. To find the contribution 
to the $k$-th term, $\mathbb{E}(Z^{k})$ of the series that defines the quenched free energy, let us define 
the operator $D_{ij}(x-y)$ in field space. We have 

\begin{equation}
D_{ij}(x-y)= \delta^{d}(x-y)\Biggl(\biggl(-c\frac{d^{\,2}}{dx^{2}}+\omega^{2}
                                    +\beta u\biggr)\delta_{ij}-\beta u\Biggr).
\label{saddle-point12}
\end{equation}

Within this definition, we can write $H_{eff}(\varphi_{i},k;\omega)$ as 

\begin{equation}
  H_{eff}(\varphi_{i},k;\omega)=\frac{1}{2}\int_{0}^{L}dx\sum_{i,j=1}^{k}\varphi_{i}(x)
                                                              D_{ij}(x-y)\varphi_{j}(x). 
\end{equation} 

Therefore, $\mathbb{E}(Z^{k})$ reads 

\begin{equation}
\mathbb{E}(Z^{k})=\prod_{i=1}^{k}\int_{\varphi_{i}(0)=0}^{\varphi_{i}(L)=y}
[d\varphi_{i}]\,\exp{\biggl(-\frac{\beta}{2}\sum_{i,j=1}^{k}\int_{0}^{L}\, dx\,\int_{0}^{L}\,dy
\,\,\varphi_{i}(x)D_{ij}(x-y)\varphi_{j}(y) \biggr)}.
\label{aa112}
\end{equation}

Substituting the above equation into Equation (\ref{serie}) we obtain the average free energy of the system.
A point that deserves be emphasized is the fact that the number of terms in the series that represents the 
average free energy can be finite. For instance, we can use the saddle-point equation to find a bound 
for $k$. Therefore, let us discus the saddle-point equations of the model. For each moment of the 
partition function, the saddle-point equations are

\begin{equation}
\biggl(-c\frac{d^{\,2}}{dx^{2}}+\omega^{2}+\beta u\biggr)\varphi_{i}(x)=
                                                                 \beta u\sum_{j=1}^{k}\varphi_{j}(x).
\label{saddle-point1}
\end{equation}

In each integer moment of the partition function, we must have $\varphi_{i}(x)=\varphi(x)$. This is the 
unique solution for the problem of the structure in field space in each moment of the partition function.  
For~ equal fields, the saddle-point equation becomes

\begin{equation}
\biggl(-c\frac{d^{\,2}}{dx^{2}}+\omega^{2}+\beta u(1-k)\biggr)\varphi(x)=0.
\label{saddle-point2}
\end{equation}     

The condition $\omega^{2}+(1-k)\beta u\geq 0$ must be satisfied to have a physical theory. 
Consider a generic term of the series given by Equation (\ref{serie}) with a moment of the partition function 
given by $\mathbb{E}\,({Z^{\,l}})$. Defining $k_{c}$ as

\begin{equation}
k_{c}=\left\lfloor\frac{ \omega^{2}}{\beta u}+1\right\rfloor,
\end{equation}
where $\left\lfloor x\right\rfloor$ means the integer part of $x$, the structure of the fields in each 
moment of the partition function is given by 

\begin{equation}
\begin{cases}
\varphi_{i}^{(l)}(x)=\varphi(x), \,\,\,\hfill\hbox{for $l=1,2,...,k_{c}$,}
\vspace{0.3cm}\\
\varphi_{i}^{(l)}(x)=0, \quad \,\,\,\,\,\hbox{for $l>k_{c}$}.
\end{cases} \label{RSB123}
\end{equation}

Only in the high-temperature limit, $(\beta\rightarrow 0)$, all the moments of the partition functions 
contribute to the average free energy. For finite temperature, we must have only a finite number of terms 
in the series representation to the average free energy. In this case, using Equation (\ref{RSB123}) the 
average free energy is given by

\begin{equation}
F_{q}=\frac{1}{\beta}\sum_{k=1}^{k_{c}}\frac{(-1)^{k}}{k!k}\,\mathbb{E}\,({Z^{\,k})}.
\label{KMener}
\end{equation} 

For $\omega\neq 0$ and large $k_{c}=N$ we have the large-N approximation for a Gaussian field theory.
It~ is interesting to point out that the limit $\omega\rightarrow\, 0$, we must have $k_{c}=1$. The system 
is described by a field theory where the dimension of the order parameter is one. In the next section, we 
discuss another Gaussian model defined in the continuous limit and calculate its wandering exponent.

\section{Field Theory for Interfaces in Random Media} \label{sec:IV}

In this section, we study an interface.  We consider a $d$-dimensional manifold with internal points, 
$x \in \mathbb {R}^{d}$, embedded in an external $D$-dimensional space with position vector 
$\vec{r}(x) \in \mathbb {R}^{D}$. We are considering  a $d$-dimensional manifold in a $D=d+N$ dimensional 
space. For oriented manifolds, we can describe the system in terms of the set of transverse coordinates, 
where $N$ is the number of transverse dimensions. We are interested in the case $N=1$, so we have an 
interface in a quenched random potential and $D=d+1$. The Hamiltonian of the domain wall can be written as  

\begin{equation}
H(\varphi,v)=\int d^{\,d}x\Biggl[\frac{\sigma}{2}|\nabla\varphi(x)|^{2}+v\bigl(\varphi(x),x\bigr)\Biggr],
\label{hamiltonian2}
\end{equation} 
where $\sigma$ is the domain wall stiffness and $v(\varphi(x),x)$ is the quenched random potential 
of the model~ \cite{mezard-parisi1,mezard-parisi3}.~ Following Mezard and Parisi and also 
Cugliandolo {et al.} \cite{mezard-parisi2,doussal}, let us introduce a 
$\frac{1}{2}\omega^{2}\varphi^{2}(x)$~ contribution,~ which constrain the manifold to fluctuate in a 
restricted volume of the embedding space. The regularized Hamiltonian becomes

\begin{equation}
H(\varphi,v)=\frac{1}{2}\int d^{\,d}x\Biggl[\varphi(x)\bigl(-\sigma\Delta
              +\omega^{2}\bigr)\varphi(x)+v\bigl(\varphi(x),x\bigr)\Biggr].
\label{hamiltonian3}
\end{equation}

The partition function of the model is given by

\begin{equation}
 Z(v)=\int[d \varphi]\exp{\bigl(-\beta H(v,\varphi)\bigr)},
\label{pf1}
\end{equation}  
where $[d \varphi]$ is a functional measure. We are assuming that the probability distribution associated 
to the random potential has zero mean

\begin{equation}
\mathbb{E}\bigl(v(\varphi,x)\bigr)=0
\label{correlation11}
\end{equation} 
and correlator 

\begin{equation}
\mathbb{E}\bigl(v(\varphi,x)v(\varphi',x')\bigr)=2V(\varphi-\varphi')\delta^{d}(x-x'),
\label{correlation22}
\end{equation} 
where again, the $\mathbb{E}(...)$ means that we are taking the average over all the realizations of the 
quenched random potential. Since we are assuming that the system has a quenched random potential, the~ 
average free energy is defined as

\begin{equation}
 F_{q}=-\frac{1}{\beta}\int[dv]P(v)\log Z(v).
\label{fe1}
\end{equation}      

With Equation (\ref{serie}) in hands, we have to compute $k$-th  moment of the partition function, 
$\mathbb{E}(Z^{k})$. After integrating over the disorder, we get

\begin{equation}
\mathbb{E}(Z^{k})=\int\,\prod_{i=1}^{k}[d\varphi_{i}]\,\exp\bigl(-\beta H_{eff}(\varphi_{i},k)\bigr),
\label{aa111}
\end{equation}
where 

\begin{equation}
H_{eff}(\varphi_{i},k)=\frac{1}{2}\int\,d^{\,d}x\Biggl[\sum_{i=1}^{k}\varphi_{i}(x)\bigl(
-\sigma\Delta+\omega^{2}\bigr)\varphi_{i}(x)-\beta\sum_{i,j=1}^{k}V\bigl(\varphi_{i}(x)-
\varphi_{j}(x)\bigr)\Bigg].
\label{heff11}
\end{equation}

Again,  as in the case of the polymers, to proceed we must use some model for 
$V(\varphi_{i}-\varphi_{j})$. Following Balents and Fisher \cite{balents-fisher}, we consider that 
$V(\varphi_{i}-\varphi_{j})$ can be written as

\begin{equation}
V(\varphi_{i}-\varphi_{j})=\sum_{m}\frac{1}{m!}V_{m}(\varphi_{i}-\varphi_{j})^{2m}.
\label{re1}
\end{equation} 

Now, let us discuss the model going beyond the Gaussian approximation. Assuming that that 
$V_{1}>0$, $V_{2}>0$ and $V_{m}=0$ for $m\geq 3$. The potential $V(\varphi_{i}-\varphi_{j})$ reads

\begin{equation*}
 V(\varphi_{i}-\varphi_{j})=V_{0}-\frac{1}{2}u_{1}(\varphi_{i}-\varphi_{j})^{2}
 -\frac{1}{4}u_{2}(\varphi_{i}-\varphi_{j})^{4}.
\label{referee1}
\end{equation*}

In this case, the $k$-th  moment of the partition function is given by
\begin{equation}
\mathbb{E}(Z^{k})=\int\,\prod_{i=1}^{k}[d\varphi_{i}]\,\exp\bigl(-\beta H_{eff}(\varphi_{i},k)\bigr),
\label{referee2}
\end{equation}
where the effective Hamiltonian can be written as 

\begin{equation}
 H_{eff}=H^{(0)}_{eff}+H^{(1)}_{eff}.
\label{referee3}
\end{equation}  

In the above equation, the Gaussian contribution is given by

\begin{equation}
 H^{(0)}_{eff}=\frac{1}{2}\sum_{i,j=1}^{k}\int\,d^{\,d}x\,\,\varphi_{i}(x)\biggl[\bigl(-\sigma\Delta
 +\omega^{2}_{0}+\beta u_{1}\bigr)\delta_{ij}-\beta u_{1}\biggr]\varphi_{j}(x)
\label{referee4}
\end{equation}
and the non-Gaussian contribution is

\begin{equation}
 H^{(1)}_{eff}=\frac{\beta u_{2}}{2}\sum_{i,j=1}^{k}\int d^{\,d}x\biggl[\frac{1}{4}\varphi_{i}^{4}(x)
 +\frac{1}{4}\varphi_{j}^{4}(x)-\varphi_{i}^{3}(x)\varphi_{j}(x)
 +\frac{3}{2}\varphi_{i}^{2}(x)\varphi_{j}^{2}(x)-\varphi_{i}(x)\varphi_{j}^{3}(x)\biggr].
\label{referee5}				
\end{equation}

We are interested in studying the structure of the field space. Using the saddle-point equations and 
assuming the symmetry ansatz, $\varphi_{i}(x)=\varphi(x)$ for all fields in each moment of the partition 
function, we have that the saddle-point equation reads

\begin{equation}
\Bigl(-\sigma\Delta\,+\omega_{0}^{2}+(1-k)\beta u_{1}\Bigr)\varphi(x)=0.
\label{sp}
\end{equation}

The condition $\omega_{0}^{2}+(1-k)\beta u_{1}\geq 0$ must be satisfied to have a physical theory. Let us 
define $k_{c}=\left\lfloor\frac{\omega_{0}^{2}}{\beta u_{1}}+1\right\rfloor$ and considering again a 
generic term of the series, given by Equation (\ref{serie}), with the moment of the  partition function, 
$\mathbb{E}\,({Z^{\,l})}$, the only choice in the field space is given by 

\begin{equation}
\begin{cases}
\varphi_{i}^{(l)}(x)=\varphi(x) \,\,\,\,\,\hfill\hbox{for $l=1,2,...,k_{c}$}
\vspace{0.3cm}\\
\varphi_{i}^{(l)}(x)=0 \quad \,\,\,\,\,\hbox{for $l>k_{c}$},
\end{cases} \label{RSB1}
\end{equation}
the average free energy becomes

\begin{equation}
F_{q}=\frac{1}{\beta}\sum_{k=1}^{k_{c}}\frac{(-1)^{k}}{k!\,k}\,\mathbb{E}\,(Z^{\,k}),
\label{KMenergy12}
\end{equation}
where $\mathbb{E}\,(Z^{\,k_{c}})$ is given by Equation (\ref{referee2}) and  the effective Hamiltonian, 
by Equations (\ref{referee3})--(\ref{referee5}). With~ the choice of the 
field space we obtain that the effective Hamiltonian can be written in the simple form 

\begin{equation}
H_{eff}(\varphi_{i},k)=\frac{1}{2}\sum_{i,j=1}^{k}\int\,d^{\,d}x\int\,d^{\,d}y\,\,
                                           \varphi_{i}(x)D_{ij}(x-y)\varphi_{j}(y),
\label{heff222}
\end{equation}   
where for simplicity we are using $u_{1}=u$ and $D_{ij}(x-y)$ reads

\begin{equation}
D_{ij}(x-y)=\biggl(\bigl(-\sigma\Delta+\omega^{2}+\beta u\bigr)\delta_{ij}-\beta u\biggr)\delta^{d}(x-y).
\label{heff223}
\end{equation}

Therefore, in Equation (\ref{KMenergy12}), the quantity $\mathbb{E}\,(Z^{k})$ is given by

\begin{equation}
\mathbb{E}\,(Z^{k})=\int\,\prod_{i=1}^{k}
[d\varphi_{i}]\,\exp\biggl[- \frac{\beta}{2}\sum_{i,j=1}^{k}\int\,d^{\,d}x\int\,d^{\,d}y\,\varphi_{i}(x)
D_{ij}(x-y)\varphi_{j}(y)\biggr].
\label{aa1111}
\end{equation}

In the symmetric ansatz framework, all the series that represents the average free energy 
can be viewed as an Euclidean field theory for a $k$-component scalar field. Defining the $k$-vector field 
$\Phi(x)$ with the components $\varphi_{1}(x),\varphi_{2}(x),...,\varphi_{k}(x)$, we can write the 
effective Hamiltonian as

\begin{equation}
H_{eff}(\Phi;k)=\frac{1}{2}\int\,d^{\,d}x\int\,d^{\,d}y\,\, \Phi^{T}(x)D(x-y;k)\Phi(y),
\label{nn1}
\end{equation}  
where $\Phi^{T}(x)$ stands for the transpose of the $k$-vector $\Phi(x)$.
In view of Equations (\ref{heff223}) and (\ref{aa1111}), the~ kernel $D(x-y;k)$ is

\begin{equation}
D(x-y;k)=\delta^{d}(x-y)\biggl(\bigl(-\sigma\Delta+\omega^{2}+\beta u\bigr){\mathbb{I}_{k}}
-\beta u\,\mathbb{M}_{k}\biggr),
\label{heff2233}
\end{equation}
where ${\mathbb {I}}_{k}$ is the $k$-dimensional identity matrix and ${\mathbb{M}}_{k}$ is the square
$k$-dimensional matrix with all elements 1 \cite{duplantier}.

Our aim now is to study the two-point correlation function of the Euclidean field theory for a $k$-
component scalar field. Performing a Fourier transform we get

\begin{equation}
H_{eff}(\varphi_{i},k)=\frac{1}{2}\sum_{i,j=1}^{k}
%\sum_{j=1}^{k}
\int\,\frac{d^{\,d}p}{(2\pi)^{d}}\,\,\varphi_{i}(p)\bigl[G_{0}
\bigr]_{ij}^{-1}(p)\varphi_{j}(-p),
\label{aa13}
\end{equation}
where $\bigl[G_{0}\bigr]_{ij}^{-1}$ is the inverse of the two-point correlation function,

\begin{equation*}
\bigl[G_{0}\bigr]_{ij}^{-1}(p)=(\sigma p^{2}+\omega^{2}+\beta u)\delta_{ij}-\beta u.
\label{aa14}
\end{equation*}

Using the projectors operators we can write the two-point correlation function $\bigl[G_{0}\bigr]_{ij}(p)$ 
as

\begin{equation}
\bigl[G_{0}\bigr]_{ij}(p)=\frac{\delta_{ij}}{(\sigma p^{2}+\omega^{2}+\beta u)}+\frac{\beta u}
{(\sigma p^{2}+\omega^{2}+\beta u)(\sigma p^{2}+\omega^{2}+\beta u(1-k))}.
\label{aa18}
\end{equation}

The first term in the right hand side of Equation (\ref{aa18}) is the bare contribution to the connected 
two-point correlation function; the second term is the contribution to the disconnected two-point 
correlation function, which becomes connected after averaging the disorder. Let us study the two-point 
correlation function. First write

\begin{equation}
[G_{0}]_{lm}(x-y)=[G_{0}]_{lm}^{(1)}(x-y)+[G_{0}]_{lm}^{(2)}(x-y;k),
\label{sf}
\end{equation}
with

\begin{equation}
[G_{0}]_{lm}^{(1)}(x-y)=\frac{\delta_{lm}}{(2\pi)^{d}}\int d^{\,d}q\frac{e^{i(x-y)q}}{\bigl(\sigma q^{2}
+\omega^{2}+\beta u\bigr)}
\label{Gxy1}
\end{equation}
and $[G_{0}]^{(2)}_{lm}(x-y;k)=[G_{0}]^{(2)}(x-y;k)$ where 

\begin{equation}
[G_{0}]^{(2)}(x-y;k)=\frac{1}{(2\pi)^{d}}\beta u\int d^{\,d}q\frac{e^{i(x-y)q}}{(\sigma q^{2}
+\omega^{2}+\beta u)(\sigma q^{2}+\omega^{2}+\beta u(1-k))}.
\end{equation}

Now we are able to discuss the functional form of $[G_{0}]_{lm}^{(1)}(x-y)$ and $[G_{0}]_{lm}^{(2)}(x-y)$.
A~ straightforward computation yields

\begin{equation}
[G_{0}]_{lm}^{(1)}(r)=\frac{\delta_{lm}}{(2\pi)^{\frac{d}{2}}r^{\frac{d-2}{2}}\sigma^{\frac{d+2}{4}}}
\bigl(\omega^{2}+\beta u\bigr)^{\frac{d-2}{4}}K_{\frac{d}{2}-1}\biggl(r\sqrt{\sigma^{-1}(\omega^{2}
+\beta u)}\biggr).
\label{G1}
\end{equation}

Also we can write that 

\begin{eqnarray} \label{G2}
[G_{0}]^{(2)}(r;k) &=& \frac{1}{(2\pi)^{\frac{d}{2}}r^{\frac{d-2}{2}}\sigma^{\frac{d+2}{4}}k}
\left[-\bigl(\omega^{2}+\beta u\bigr)^{\frac{d-2}{4}}K_{\frac{d}{2}-1}\left(r\sqrt{\sigma^{-1}(\omega^{2}
+\beta u)}\right)\right.  \nonumber \\ 
  & & \left.+\bigl(\omega^{2}+\beta u(1-k)\bigr)^{\frac{d-2}{4}}
   K_{\frac{d}{2}-1}\left(r\sqrt{\sigma^{-1}(\omega^{2}+\beta u(1-k))}\right)\right] .
\end{eqnarray}

In the case which we are interested in, one must take the limit of $\omega\rightarrow 0$, since 
$\omega_{0}^{2}+(1-k)\beta u_{1}\geq 0$, only the contribution from $k=1$ survives. From Equations (\ref{G1}) 
and (\ref{G2}) we have 

\begin{equation}
[G_{0}](r)=\frac{1}{4(\pi \sigma)^{\frac{d}{2}}}\Gamma\biggl(\frac{d-2}{2}\biggr)\frac{1}{r^{d-2}}.
\end{equation}

We can introduce a normalized generating functional with the normalization factor $(det\,'\, D)^{-1/2}$ 
where the prime sign means that the contribution of the zero mode must be omitted. The wandering exponent 
describes the growth of the transverse fluctuations of the manifold as function of the distances. In the 
limit $\omega\rightarrow 0$, we obtain from Equation (\ref{aa18}) the wandering exponent given by 
$\xi=\frac{4-d}{2}$ as was discussed by Parisi and M\'ezard in Ref. \cite{mezard-parisi2}, discussing the 
replica symmetric solution.

\section{Conclusions} \label{sec:V}
For quenched disorder, we are mainly interested in averaging the free energy over the disorder, which 
amounts to averaging the logarithm of the partition function. The standard replica method is a powerful 
tool used to calculate the free energy of systems with quenched disorder.
In this paper, we  computed the average free energy of directed polymer and a fluctuating interface in the 
presence of a quenched disorder. To find the average free energy for both systems, we~ define a 
distributional zeta-function. The derivative of this function at the origin yields the average free energy 
of the underlying system. The average free energy of a system with quenched disorder is represented by a 
series in which all the moments of the partition functions contribute. 
In the case of a polymer and an interface in a quenched random potential, we were able to discuss the 
field theory generated by each term of the series that defines the average free energy. Each term of the 
series that represents the average free energy is an Euclidean field theory for a $k$-component scalar 
field. As~ an application of the distributional zeta-function method, we calculate the wandering exponent 
in a generic $d$-dimensional manifold. We obtain that the wandering exponent given by $\xi=\frac{4-d}{2}$. 
This~ result was discussed by Parisi and M\'ezard in Ref. \cite{mezard-parisi2} in the replica symmetry 
scenario.

\section{Acknowledgments}
We would like to acknowledge Benar F. Svaiter, Carlos A. D. Zarro and Gabriel Menezes for the discussions. 
This research was funded by Conselho Nacional de Desenvolvimento Cientifico e Tecnológico do
Brazil-(CNPq), 301751/2019-6(NFS).


\begin{thebibliography}{999}
\bibitem{membranas} Nelson, D.; Piran T.; 
Weinberg, S. (Eds.) {\em Statistical Mechanics of Membranes and Surfaces};  World Scientific: Singapore, 2004.
\bibitem{halpin} Halpin-Healy, T.; Zhang, Y. C. Kinetic roughening phenomena, stochastic growth, directed 
polymers and all that. Aspects of multidisciplinary statistical mechanics. {\em Phys. Rep.} {\bf 1995}, 
{\em 254}, 215--414.
\bibitem{fredrickson} Fredrickson, G. H. {\em The Equilibrium Theory of Inhomogeneous Polymers}, Oxford 
University Press: Oxford, UK, 2006.
\bibitem{kacokatso} Kawkatsu, T. {\em Statistical Physics of Polymers}; Springer: 
Berlin/Heidelberg, Germany, 2004.  
\bibitem{deGennes} De Gennes, P. G. Exponents for the excluded volume problem as derived by the 
Wilson method. {\em Phys.~ Lett.~ A} {\bf 1972}, {\em 38}, 339--340.
\bibitem{des} des Cloizeaux, J. The Lagrangian theory of polymer solutions at intermediate concentrations. 
{\em Phys. France} {\bf 1975}, {\em 36}, 281--291. 
\bibitem{witten} Sch\"afer, L.; Witten,  T.A., Jr. Renormalized field theory of polymer solutions. I. 
scaling laws. {\em J. Chem. Phys.} {\bf 1977}, {\em 66}, 2121. 
\bibitem{vilgis} Vilgis, T.A. Polymer theory: path integrals and scaling. {\em Phys. Rep.} {\bf 2000}, 
{\em 336}, 167--254.
\bibitem{craig} Craig, A.; Terentjev, E.M.; Edwards, S.F. Polymer localization in random potential 
{\em Physica A} {\bf 2007}, {\em 384}, 150--164.
\bibitem{duo} Kardar, M.; Nelson, D.R. $\epsilon$ expansions for crumpled manifolds. {\em Phys. Rev. 
Lett.} {\bf 1987}, {\em 58}, 1289.
\bibitem{du0}  Kardar, M.; Nelson, D.R. Statistical mechanics of self-avoiding tethered manifolds. 
{\em Phys. Rev. A} {\bf 1988}, {\em 38}, 966.
\bibitem{du1} Duplantier, B. Interaction of crumpled manifolds with Euclidean elements. {\em Phys. Rev. 
Lett.} {\bf 1989}, {\em 62}, 2337.
\bibitem{nat} Nattermann T.; Rujan, P. Random field and others systems dominated by disorder fluctuations. 
{\em Int. J. Mod. Phys. B} {\bf 1989}, {\em 3}, 1597–-1654.
\bibitem{mezard-parisi1} M\'ezard, M.; Parisi, G. Interfaces in a random medium and replica symmetry 
breaking. {\em J. Phys. A Math. Gen.} {\bf 1998}, {\em 23}, L1229.
\bibitem{mezard-parisi3} M\'ezard, M.; Parisi, G. Manifolds in random media: two extreme cases. {\em J. 
Phys. I France} {\bf 1992}, {\em 2}, 2231--2242.
\bibitem{edwards} Edwards, S.F.; Anderson, P.W. Theory of spin glasses. {\em J. Phys. F} {\bf 1975}, 
{\em 5}, 965--974.
\bibitem{livro1} M\'ezard, M.; Parisi, G.; Virasoro, M. {\em Spin-Glass Theory and Beyond}; World 
Scientific: Singapore, 1987.
\bibitem{parisi1} Parisi, G. An introduction to the statistical mechanics of amorphous systems. In {\em 
Field Theory, Disorder and Simulations}; Word Scientific: Singapore, 1992.
\bibitem{livro3} Dotsenko, V. {\em Introduction to the Replica Theory in Disordered Statistical 
Systems}; Cambridge University Press: Cambridge, UK,  2001.
\bibitem{livro4} De Dominicis, C.; Giardina, I. {\em Random Fields and Spin Glass}; Cambridge University 
Press: Cambridge, UK, 2006.
\bibitem{livromez} M\'ezard, M.; Montarini, A. {\em Information, Physics and Computation}: Oxford 
University Press: New York, NY, USA, 2008.
\bibitem{errado0} van Hemmen, J.L.; Palmer, R.G. The replica method and solvable spin glass model. 
{\em J. Phys. A} {\bf 1979}, {\em 12 }, 563.
\bibitem{errado1} Verbaarschot, J.J.M.; Zirnbauer, M.R. Critique of the replica trick. {\em J. Phys. A
Math. Gen.} {\bf 1985}, {\em 17}, 1093.
\bibitem{errado2} Zirnbauer, M.R. Another critique of the replica trick. {\em arXiv} 
{\bf 1999}, arXiv:cond-mat9903338.
\bibitem{errado3} Kanzieper, E. On exact integrability of replica field theories in 0 dimensions: 
non-Hermitean disordered Hamiltonians. {\em HAIT J. Sci. Eng.} {\bf 2004, }{\em 1}, 
101--129.
\bibitem{FTdisorsys} Parisi, G. Field theory and the physics of disordered systems.  { In Proceedings of the  Quarks, 
Strings and the Cosmos---Hector Rubinstein Memorial Symposium}, Stockholm, Sweden, 9--11 August { 2010}; Volume  
{ 109}.
\bibitem{sourlas} Fytas, N.G.; Martin-Maior, V.; Picco, M.; Sourlas, N. Restoration of dimensional 
reduction in the random-field Ising model at five dimensions. {\em Phys. Rev. E} {\bf 2017}, {\em 95},
042117.
\bibitem{ma} Ma, S.-k. {\em Modern Theory of Critical Phenomena}; Perseus Publishing: Cambridge, UK,  2000.
\bibitem{gaspari} Gaspari, G.; Rudnick, J. $n$-vector model in the limit $n\rightarrow0$ and the 
statistics of linear polymer systems: A~ Ginzburg-Landau theory. {\em Phys. Rev. B} {\bf 1986}, 
{\em 33}, 3295.
%\bibitem{schafer} Sch\"{a}fer, L., Witten Jr., T.A. Renormalized field theory of polymer solutions. I. 
%Scaling laws. {\em J. Chem. Phys.} {\em 66}, 2121, 1977.
\bibitem{mezard-parisi2} M\'ezard M.; Parisi, G. Replica field theory for random manifold. {\em J. Phys. 
I France} {1991}, {\em 1}, 809--836.
\bibitem{svaiter1} Svaiter, B.F.; Svaiter, N.F. The distributional zeta-function in disordered field 
theory. {\em Int. J. Mod. Phys. A} {\bf2016}, {\em 31}, 1650144.
\bibitem{svaiter2} Svaiter, B.F.; Svaiter, N.F. Disordered field theory in d=0 and distributional 
zeta-function. {\em arXiv} {\bf 2016}, arXiv:math-phys1606.04854.
\bibitem{novo} Acosta Diaz, R.; Svaiter, N.F.; Menezes, G.; Zarro, C.A.D. Spontaneous symmetry breaking in 
replica field theory. {\em Phys. Rev. D} {\bf 2017},  {\em 96}, 065012.
\bibitem{gastao} Acosta Diaz, R.; Svaiter, N.F.; Krein, G.; Zarro, C.A.D. Disordered $\lambda\phi^{4}+
\rho\phi^{6}$ Landau-Ginzburg model. {\em Phys.~ Rev.~ D} {\bf 2018}, {\em 97}, 065017.
\bibitem{novo1} Acosta Diaz, R.; Krein, G.; Saldivar, A.; Svaiter, N.F.; Zarro, C.A.D. Disordered 
Bose–Einstein condensate in hard walls trap. {\em J. Phys. A} {\bf 2019}, {\em 52}, 445401.
\bibitem{novo2} Soares, M.S.; Svaiter, N.F.; Zarro, C.A.D. Multiplicative noise in Euclidean 
Schwarzschild manifold. {\em Class.~ Quant.~ Grav.} {\bf 2020}, {\em 37}, 065024.
\bibitem{larkin}  Larkin, A. Effect of inhomogeneities on the structure of the mixed state of 
superconductors. {\em Sov. Phys. JETP} {\bf 1970}, {\em 31}, 784.
\bibitem{huse} Huse, D.A.; Henley, C.L. Pinning and roughening of domain walls in Ising systems due to 
random impurities. {\em Phys. Rev. Lett.} {\bf 1985}, {\em 54}, 2708.
\bibitem{kardar} Kardar, M. Replica Bethe ansatz studies of two-dimensional interfaces with quenched 
random impurities. {\em Nucl. Phys. B} {\bf 1987}, {\em 290}, 582.
\bibitem{garel} Monthus, C.; Garel, T. Directed polymers and interfaces in random media: Free-energy 
optimization via confinement in a wandering tube. {\em Phys. Rev. E} {\bf 2004}, {\em 69}, 061112.
\bibitem{digkb} Dotsenko, V.S.; Ioffe, L.B.; Gerhkerbein, V.B.; Korshunov  S.E.; Blatter, G. Joint 
free-energy distribution in the random directed polymer problem. {\em Phys. Rev. Lett.} {\bf 2008}, 
{\em 100}, 050601.
\bibitem{dggb} Dotsenko, V.S.; Gerhkerbein, V.B.; Korshunov, S.E.; Blatter, G. Free-energy distribution 
functions for the randomly forced directed polymer. {\em Phys. Rev. B} {\bf 2010}, {\em 82}, 174201.
\bibitem{periodic} Acosta Diaz, R.; Svaiter, N.F. Finite-size effects in disordered $\lambda\phi^{4}$ 
model. {\em Int. J. Mod. Phys. B} {\bf 2016}, {\em 30}, 1650207.
\bibitem{riem2} Ingham, A.E. {\em The Distribution of Prime Numbers}; Cambridge University Press:
Cambridge, UK, 1990.
\bibitem{primes0} Menezes, G.; Svaiter, N.F. Quantum field theories and prime numbers spectrum. 
{\em arXiv} {\bf 2011}, arXiv:1211.5198.
\bibitem{primes} Menezes, G.; Svaiter, B.F.; Svaiter, N.F. Riemann zeta zeros and prime number spectra in 
quantum field theory. {\em Int. J. Mod. Phys. A} {\bf 2013}, {\em 28}, 1350128.
\bibitem{voros} Voros, A. {\em Zeta Functions over Zeros of Zeta Functions}; Springer:  {Berlin/Heidelberg, Germany,} 
, 2010.
\bibitem{nami1} Svaiter, N.F.; Svaiter, B.F. Casimir effect in a $d$‐dimensional flat space‐time and the 
cut‐off method. {\em J.~ Math.~ Phys.} {1991}, {\em 32}, 175.
\bibitem{nami2} Svaiter, N.F.; Svaiter, B.F. The analytic regularization zeta function method and the 
cut-off method in the Casimir effect. {\em J. Phys. A} {\bf 1992}, {\em 25}, 979.
\bibitem{nami3} Svaiter, B.F.; Svaiter, N.F. Zero point energy and analytic regularizations. 
{\em Phys. Rev. D} {\bf 1993}, {\em 47}, 4581.
\bibitem{doussal} Cugliandolo, L.F.; Kurchan, J.; Le Doussal, P. Large Time out-of-equilibrium dynamics of 
a manifold in a random potential. {\em  Phys. Rev. Lett.} {\bf 1996}, {\em 76}, 2390.
\bibitem{balents-fisher} Balents, L.; Fisher, D.S. Large-$N$ expansion of $(4-\epsilon)$-dimensional 
oriented manifolds in random media. {\em Phys. Rev. B} {\bf 1993}, {\em 48}, 5949.
\bibitem{duplantier} Duplantier, B. Statistical mechanics of polymer networks of any topology. {\em 
J. Stat. Phys.} {1989}, {\em 54}, 581–-680.

% Reference 1
%\bibitem[Author1(year)]{ref-journal}
%Author1, T. The title of the cited article. {\em Journal Abbreviation} {\bf 2008}, {\em 10}, 142--149.

\end{thebibliography}
\end{document}